\documentstyle[epsfig]{aipproc}
\begin{document}

\author{I.I. Mazin and D. J. Singh}
\title{Weighted Density Functionals for Ferroelectric Materials}
\maketitle

\begin{abstract}
The weighted density approximation, its implementation and its application
to ferroelectric materials is discussed. Calculations are presented for
several perovskite oxides and related materials. In general the weighted
density approximation is found to be superior to either the local density or
generalized gradient approximation for the ground state. Electronic
structures are little changed. The linear response of the weighted density
approximation is calculated for the homogeneous electron gas, and found to
be improved relative to the local density result, but not in full agreement
with existing Monte Carlo data. It is shown that the agreement can be
further improved by a simple modification. Calculations of the
ferroelectric soft mode in KNbO$_3$ suggest that the low temperature
distortion is approximately 20\% smaller than indicated by existing
experiments.
\end{abstract}

\address{Complex Systems Theory Branch, Naval Research Laboratory,
Washington, DC 20375-5320}

\section{Introduction}

Piezoelectric, ferroelectric and related perovskites are both
technologically important and physically interesting systems and, as such,
have been the subject of considerable recent interest. Several groups have
performed first principles calculations based on density functional theory.
These calculations, primarily based on the local density approximation (LDA)
have elucidated important aspects of the underlying physics of these
materials as well as providing quantitative information about phonons,
polarizations, crystal structure, elastic and other properties. \cite{proc97}

Two of the many themes that have emerged are (1) Ferroelectric and
piezoelectric properties are due to a delicate balance between large
competing interactions, \cite{c92,c92a} so very high accuracy is required for
predictive calculations and (2) The lattice instabilities behind these
phenomena are extraordinarily sensitive to volume. In fact, in important
materials, like KNbO$_{3}$ \cite{s92,p93,p95,k95,w97,s95,s97} and BaTiO$_{3}$
\cite{c92,c92a,ck90} ferroelectricity, while predicted at the experimental
volume, is absent, or at least strongly suppressed, at the predicted LDA
volume.

Generally excellent agreement with experimental knowledge has been obtained
when calculations are performed at the experimental volume. This includes
even transition temperatures, polarizations and transition temperatures \cite
{proc97}. In antiferroelectric PbZrO$_{3}$, LDA calculations \cite{singh-pb}
at the experimental volume yielded a crystal structure that differed from
existing experimental measurements. However, the LDA structure has now been
confirmed by two independent neutron investigations, \cite{egami,fujishita}
providing yet another demonstration of the utility of the first principles
approach. On the other hand, even at the experimental volume, the LDA
consistently overestimates the dielectric constant, $\varepsilon _{\infty }$%
. \cite{dalcorso,resta1,resta2} Exploration of novel systems that are of
necessity less well studied experimentally than current materials will
require an ability to calculate properties independent of experimental
measurements of the volume. This may also be important for determination of
piezoelectric properties in materials like PbTiO$_{3}$ where the LDA gives
an incorrect prediction of the strain associated with the deviation of the $%
c/a$ ratio from unity even with the volume constrained. \cite{ss} This
raises the question of whether there is a practical way of correcting the
LDA errors, in order to construct a more predictive first principles
approach for these materials.

The LDA errors can sometimes be trivially fixed. For example, if the
experimental volume is known, the calculations may be done constrained to
this volume. In some cases, e.g. KNbO$_3$, simple modification of the LDA
functional through use of the Wigner form rather than an electron gas-like
form, \cite{k95,w97} or the use of generalized gradient approximations (GGA)
to density functional theory are sufficient to correct the volume (although
the dielectric constant remains too high). However, in general these
approaches do not solve the LDA problem. For example, in BaTiO$_3$ the GGA
overcorrects the LDA volume, so that the error is as large but in the
opposite direction. \cite{s95} The implication is that a more sophisticated,
presumably non-local, density functional is needed to fully circumvent the
problems noted above with the LDA.

The first efforts at developing practical non-local functionals date from
the 1970's when the average density approximation\cite{olle1} (ADA) and
weighted density approximation\cite{olle2,AG,GJL} (WDA) were proposed.
However, over most of the intervening period the field has been relatively
dormant, in part because of the success of the simpler LDA and GGA schemes
and in part because it was widely thought that such schemes could not be
implemented in a computationally tractable fashion. However, at least for
the WDA, computationally efficient algorithms are now known \cite
{s97,Singh,tet,char} and benchmark calculations have been reported. The ADA
has attracted less interest.

In the cases that have been studied ground state properties of solids are
generally improved over the LDA. \cite{char1} These include tests for
several simple elements and compounds and KNbO$_3$. Significantly, this
latter test showed that the WDA predicts an equilibrium volume for KNbO$_3$
in almost perfect agreement with experiment. \cite{s97} Unlike the GGA,
which also greatly improves the equilibrium volume of KNbO$_3$, the WDA is a
truly non-local density functional, in that the exchange correlation
potential at a point {\bf r} incorporates information about the charge
density $n({\bf r}^{\prime})$ over a finite region of space.

Here, following a brief overview of the method, we report investigations of
a range of perovskite and related oxides within the WDA. These show much
improved volumes relative to either the LDA or GGA. Phonon frequencies and
the ferroelectric soft mode in KNbO$_{3}$ are calculated within the WDA and
compared with LDA results and experiment. An analytical expression for the
linear response of the uniform electron gas in the WDA is derived.

\section{The Weighted Density Approximation}

The WDA and the ADA are both based on the general expression for the
exchange correlation energy of a general electron gas $E_{xc},$ in density
functional theoryi (DFT). 
\begin{equation}
E_{xc}=\frac{e^{2}}{2}\int \frac{n({\bf r})n({\bf r}^{\prime })}{|{\bf r-r}%
^{\prime }|}G({\bf r,r}^{\prime })\{n({\bf r)}\}d{\bf r}d{\bf r}^{\prime },
\label{exc}
\end{equation}
where the function $G({\bf r,r}^{\prime })$ is also a functional of the
total electronic density $n{\bf (r).}$ A rigorous expression for $G$ can be
derived\cite{IEG} in terms of coupling constant averaged pair correlation
function: 
\begin{equation}
G({\bf r,r}^{\prime })=\int_{0}^{1}[\bar{g}({\bf r,r}^{\prime };\lambda )\{n(%
{\bf r)}\}-1]d\lambda .  \label{realG}
\end{equation}
Were the actual pair correlation function used instead of the coupling
constant averaged one, this expression would give the interaction energy of
each electron with its exchange correlation hole. The difference due to the
averaging reflects the fact that kinetic energy of interacting electrons
differs from that of a non-interacting system with the same density. It is
this difference, also known as exchange-correlation kinetic energy, which is
responsible for the high {\it q}, short wavelength behavior of the response,
and is also implicated in usual
 underestimation of semiconducting gaps in Kohn-Sham DFT.

For the uniform gas this function, $G_{0}(|{\bf r-r}^{\prime }|,n),$ is
known with high accuracy\cite{PW}, but for an arbitrary inhomogeneous
system, like occurs in a real density functional calculation for a solid, $G$
is not known and there is therefore no practical way to use this formula
without making some approximation.

The LDA instead of Eq. (\ref{exc}) uses $(e^{2}/2)\int d{\bf r}d{\bf r}%
^{\prime }n^{2}({\bf r})G_{0}[|{\bf r-r}^{\prime }|,n({\bf r)]}/|{\bf r-r}%
^{\prime }|,$ so that $E_{xc}$ becomes $E_{xc}^{LDA}=\int n({\bf r})\epsilon
_{xc}[n({\bf r)]}d{\bf r,}\ \ \epsilon _{xc}$ being the density of
exchange-correlation energy of the uniform gas. The LDA is incorrect in both
most important limits: the fully localized, {\it i.e.}, a one electron
system, and the fully delocalized limit, {\it i.e.}, homogeneous electron
gas. In the former case the LDA gives a spurious self-interaction with
energy $(e^{2}/2)\int d{\bf r}d{\bf r}^{\prime }n({\bf r})n({\bf r}^{\prime
})/|{\bf r-r}^{\prime }|+\int n({\bf r})\epsilon _{xc}[n({\bf r)]}d{\bf r},$
which is widely thought \cite{sic} to be a key problem with the LDA. In the
homogeneous limit, the LDA gives the correct exchange-correlation energy,
but the {\it changes} of this energy upon small perturbations are not
properly described; the second variation of $E_{xc}$ with density, {\it i.e.}%
, the exchange-correlation part of the dielectric response, $K_{xc}({\bf r-r}%
^{\prime })=\delta ^{2}E_{xc}/\delta n({\bf r})\delta n({\bf r}^{\prime }),$
is a delta function, which is incorrect. The Fourier transform of $K_{xc}(r)$
in LDA is independent of the wave vector. Since LDA is exact for the uniform
gas, $K_{xc}^{LDA}$ corresponds to the correct $K_{xc}$ at $q=0.$ GGAs also
give correct behavior at $q=0$, but become even worse than the LDA at high $q
$'s.

The two nonlocal expressions for $E_{xc}$, WDA and ADA, are aimed at correcting
one or the other of these two limits. The former uses the general expression
(\ref{exc}), but instead of the actual function $G$ uses a model function,
defined so that the one electron limit is honored. This begins by choosing a
generic expression for $G,$ which depends on one parameter $\bar{n},$ to be
defined later. In the original papers it was suggested that $G({\bf r,r}%
^{\prime },\bar{n})=G_{0}({\bf r,r}^{\prime },\bar{n})=\int_{0}^{1}[\bar{g}(%
{\bf r,r}^{\prime };\lambda ,\bar{n})-1]d\lambda $, where $\bar{g}$ is the
averaged pair correlation function of the homogeneous electron gas. Later it
was realized\cite{GJ} that other choices of $G$ may be better than $G_{0},$
and that there is no physical reason to prefer $G_{0}$ over many other
choices. In the WDA $\bar{n}$ is a function of ${\bf r,}$ but differs from $%
n({\bf r),}$ and is chosen so that $\int G[{\bf r,r}^{\prime },\bar{n}({\bf r%
})]n({\bf r}^{\prime }{\bf )}d{\bf r}^{\prime }=-1.$ This assures that for a
one electron system $E_{xc}$ cancels the self-interaction exactly.

Within the WDA, non-local information about the charge density is
incorporated into $E_{xc}$ both through the construction of $\bar{n}$ via
the sum rule, and through the non-local Coulomb integral.

We note that $G$ need not neccesarily be the actual pair correlation
function of the system: although Eq.(\ref{exc}) has the same functional form
as the WDA energy, the fact that $G^{WDA}$ is a function of an averaged
density $\bar{n} $, not a functional of the true density $n({\bf r})$, means
that it is possible that the best approximations for $G^{WDA}$ could be
different from the physical function $G$ defined in Eq. (\ref{realG}) even
for the uniform electron gas. However, the few reported calculations suggest
that this function, while perhaps not optimal, does yield results much
superior to the LDA. Besides, use of the homogeneous electron gas averaged
pair correlation function has an appealing simplicity and calculations of
ground state properties, below, use this ansatz. The results confirm that at
least for the structural properties considered, the WDA with this choice of $%
G$ yields results that are uniformly superior to the LDA.

\section{Ground State Properties in the WDA}

As mentioned, one of the basic problems with application of the LDA to
ferroelectric materials is the high sensitivity of ferroelectric properties
to the volume, combined with the several percent errors in predicted LDA
equilibrium volumes. One of our main goals is to determine whether the WDA
can repair this problem while retaining the good features of the LDA. We
begin by calculating the equilibrium volumes of several cubic or near cubic
oxides, namely CaO, SrO, BaO, BaTiO$_3$, KNbO$_3$ and KTaO$_3$. Some results
for KNbO$_3$ were reported previously. \cite{s97} The LDA underestimates the
cubic lattice parameters of these materials by 1-2\%, while at least in BaO
and BaTiO$_3$ GGA calculations seriously overestimate the volume. \cite{s95}

The calculations were done using a planewave basis set pseudopotential
method, \cite{pwcode1,pwcode2} with hard Troullier-Martins pseudopotentials 
\cite{tm2} and the implementation of the WDA discussed in Ref. %
\onlinecite{Singh}. The Perdew-Wang form of $G$, i.e. that from the uniform
electron gas coupling constant averaged pair correlation function was used
for the WDA. The pseudopotentials included the semi-core states as valence
states. In particular, $3s$ and $3p$ states of K, Ti and Ca, $4s$ and $4p$
states of Nb and Sr and $5s$ and $5p$ states of Ba and Ta were treated
in the electronic structure calculations, while lower lying states
were pseudized. Well converged basis sets, including planewaves up to 121
Ry, were used, except for CaO where a higher cut-off of 132 Ry was used.

One complication in the WDA is the need to use shell partitioning to avoid
unphysical exchange-correlation interactions between valence and core
electrons. It is used to prevent the core electrons from contributing to the
exchange correlation hole seen by valence electrons, since it is
unreasonable for tightly bound core electrons to dynamically screen valence
electrons. In shell partitioning, the interaction between the valence
electrons is treated with the WDA, but core-core and core-valence
interactions are treated within the LDA as discussed in Refs. %
\onlinecite{Singh,s97}. In the present calculations, the semi-core states on
the metal ions and the O $2s$ states are treated with the LDA, lower states
are pseudized with an LDA pseudopotential, and higher states are treated
with the WDA.

The LDA and WDA lattice parameters of these oxides are summarized in Table 
\ref{tab1} along with some previous results for elements. As may be seen,
the WDA lattice parameters are quite dramatically improved relative to the
LDA, including BaTiO$_3$ and BaO, for which GGA calculations yield
overcorrected volumes.

\begin{table*}
\caption{LDA, WDA and experimental lattice parameters in \AA ~for some
cubic materials. Values for the elements C, Si, Mo and V and are from
Ref. \protect\onlinecite{Singh}.
BaTiO$_3$ and KNbO$_3$ are really rhombohedral at
low temperature, but the rhombohedral strain is small. Calculations
were done for the cubic structure. Numbers in parentheses are the
percentage deviation from experiment.}
\begin{tabular*}{\linewidth}{lccc|lccc}
Material&LDA&WDA&Expt.&
Material&LDA&WDA&Expt.\\
\tableline
C\hfill &3.53 (-1.1) &3.56 (-0.3)&3.57&
Si&5.36 (-1.3)&5.40 (-0.6)&5.43\\
Mo&3.11 (-1.3)&3.14 (-0.3)&3.15&
V&2.93 (-3.0)&2.99 (-1.0)&3.02\\
CaO&4.71 (-2.0)&4.81 (-0.0)&4.81&
SrO&5.06 (-1.8)&5.16 (-0.0)&5.16\\
BaO&5.46 (-1.4)&5.56 (+0.3)&5.54&
BaTiO$_3$&3.95 (-1.2)&4.00 (-0.1)&4.00\\
KNbO$_3$&3.96 (-1.4)&4.02 (+0.1)&4.02&
KTaO$_3$&3.92 (-1.6)&3.98 (-0.2)&3.98\\
\end{tabular*}
\label{tab1}
\end{table*}

Next we turn to phonons and the ferroelectric soft mode. The above results
imply that the WDA can generally correct LDA errors in the equilibrium
volume. The question then arises as to the extent that the WDA is able to
reproduce the desirable features of the LDA with the volume corrected. LDA
calculations at the experimental lattice parameter have demonstrated very
good agreement with experimental data for phonons, crystal structures and
even derived quantities like transition temperatures. Moreover, in
ferroelectric/piezoelectric materials these quantities are governed by 
delicate balances between competing terms. In KNbO$_3$, very different
results are obtained for the ferroelectric soft mode if the lattice
parameter is varied by  1-2 percent, and in fact the WDA does change the
LDA prediction of the lattice parameter by an amount of this order. Thus one
is led to ask whether the WDA will drastically change LDA results for
phonons or the ferroelectric mode or otherwise degrade the description of
the material.

To begin addressing this issue we performed frozen phonon calculations of
the zone center transverse
modes in KNbO$_3$. Parallel LDA and WDA planewave
pseudopotential were performed as described above using a 6x6x6 special
k-point sampling of the Brillouin zone.

Calculations of atomic forces were performed for small (less than 0.05{\AA}) 
displacements of the atoms from the ideal cubic sites in a rhombohedral
symmetry consistent with the $\Gamma _{15}$ modes. A total of seven such
force calculations were performed. The $\Gamma _{15}$ dynamical matrix was
then least squares fit to these force calculations, and diagonalized to
obtain the frequencies and eigenvectors. The LDA and WDA eigenvectors were
very similar to each other and to previously reported \cite{s92,s97}
linearized augmented planewave (LAPW) \cite{LAPW} results, and are not
reproduced here. There is only a single $\Gamma _{25}$ frequency, which was
calculated separately, using force calculations with O displacements
according to this mode.

\begin{table*}
\caption{LDA and WDA $\Gamma$-point TO phonon frequencies for
cubic perovskite structure KNbO$_3$. The lattice parameter is
fixed at its experimental value, which is the same as the WDA
value.}
\begin{tabular*}{\linewidth}{ccc}
$\Gamma$-point mode \hfill 
&LDA frequencies (cm$^{-1})$&\ \ WDA frequencies (cm$^{-1}$)\\ 
\tableline
$\Gamma_{15} (unstable)$&  195 {\it i}&176 {\it i}\\
$\Gamma_{15}$&  170&143\\
$\Gamma_{15}$&  473&497\\
$\Gamma_{25}$&  247&257\\
\end{tabular*}
\label{tab2}                              
\end{table*}

Calculated LDA and WDA frequencies of the $\Gamma _{15}$ and $\Gamma _{25}$
modes in cubic perovskite structure KNbO$_{3}$ are given in Table \ref{tab2}%
. The LDA results are similar to those obtained previously by all electron
LAPW,\cite{s92} pseudopotential linear response LAPW methods\cite{k95,w97}
and planewave pseudopotential calculations for the $\Gamma _{15}$ modes. 
\cite{zhong} The agreement of the LDA results with the linear response
calculations of Yu and co-workers {\it et al.} \cite{k95,w97} is almost
exact, with a maximum deviation of 4 cm$^{-1}$, while somewhat larger
differences from the earlier all-electron results of Singh and Boyer \cite
{s92} are present. Tests \cite{c92a,s95} have shown that most of this latter
difference is due to the use of a 4x4x4 k-point mesh in the calculations of
Ref. \onlinecite{s92} leading to an underestimation of the ferroelectric
instability.  Fontana {\it et al.} \cite{fontana} have performed
measurements of phonon frequencies in KNbO$_{3}$. These should not be
uncritically compared to the present ground state calculations because the
KNbO$_{3}$ undergoes three structural transitions as the temperature is
raised from 0K, and only becomes cubic at approximately 700K. Nonetheless,
as discussed in Ref. \onlinecite{s92}, they are consistent with the LDA
results for the stable modes.

The WDA phonon frequencies are very similar to the LDA results. The main
difference is that the splitting between the ferroelectric soft mode and the
middle (lowest stable) $\Gamma_{15}$ mode is smaller in the WDA. However,
differences between the LDA and WDA frequencies are comparable to the
numerical accuracy of the calculations as estimated from the spread between
the recent results of various groups. In any case, the present results do
not provide any indication that the WDA degrades the favorable agreement of
LDA phonon frequencies with experiment in these materials although further
tests are clearly in order.

\begin{figure}[h!]
\centerline{\epsfig{file=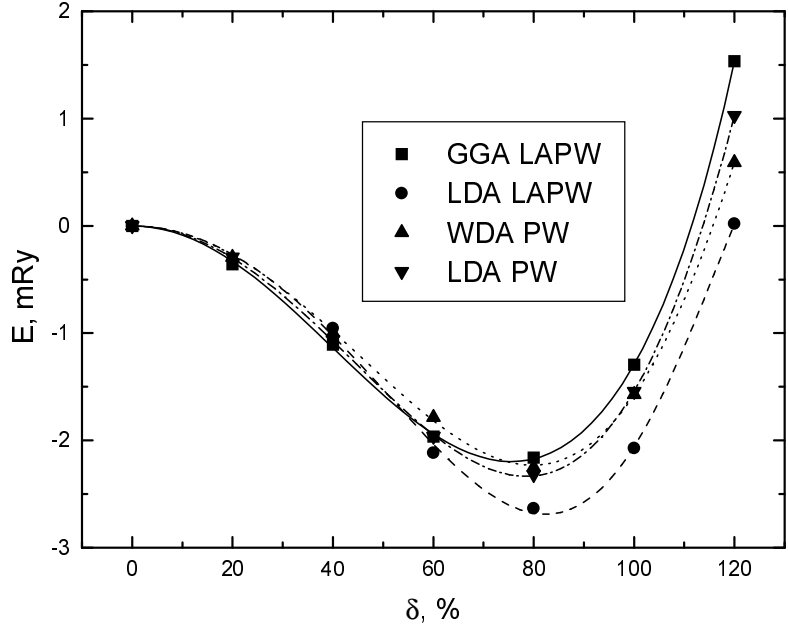,width=0.9\linewidth}}
\vspace{0.1in} \nopagebreak
\caption{Total energy as a function of ferroelectric displacement
in KNbO$_3$ within the WDA, LDA and GGA using both planewave 
pseudopotential (LDA and WDA) and LAPW approaches (LDA and GGA).
The displacement $\delta$ is in percent of
the experimental displacement of Ref.
\protect\onlinecite{hewat}. Calculations are at the experimental volume, 
which effectively coincides with the WDA and GGA volumes.}
\end{figure}

Next we turn to the ferroelectric instability. The present LDA and WDA
calculations done with a planewave pseudopotential method, and earlier LAPW
calculations both within the LDA and with GGA density functionals are in
agreement with the experimental distortion \cite{hewat} regarding the soft
mode eigenvector. However, both LDA and GGA calculations give a soft mode
amplitude that is 20\% smaller than the existing experimental value, which
was determined by powder neutron diffraction. Interestingly, both LDA and
GGA calculations\cite{s95,s97}
 for BaTiO$_3$ give a distortion that agrees very
closely, (within 5\%) with experiment. Fig. 1 shows the result of WDA
calculations of this instability for KNbO$_3$
 at the experimental volume, in terms of the
reported experimental distortion. As may be noted, there is some scatter
between the curves, which represent three different exchange correlation
potentials and two different band structure methods. However, all the
calculations are in agreement that there is an instability of order 2
mRy/f.u. and that the distortion is only 80\% of the reported experimental
distortion. In particular, the WDA does not significantly change the
distortion relative the the LDA value. Taken together the results strongly
suggest an experimental re-examination of the low temperature structure of
KNbO$_3$.

\section{Linear Response in the WDA}

As mentioned, LDA systematically overestimates the static dielectric
constant, $\varepsilon _{\infty }$. Not only is this itself an important
characteristic of ferroelectric materials, it also underscores the fact
that the dielectric response in general is not accurately reproduced, which
could lead to difficulties calculating small energy differences associated
with phonons and lattice distortions. The problem has been discussed in 
terms of 
 the differences between the real one-electron excitation spectrum and the
LDA band structure, primarily to underestimation of the band gap. An
apparent paradox here is that the exact DFT theory should reproduce the
static response functions, but not the excitation spectrum. The solution of
this paradox is that the RPA (random phase) dielectric function, which is
directly defined by the one-electron spectrum, is enhanced by the so-called
exchange-correlation local field corrections. Correspondingly, a small gap
and small local field corrections result in the same $\varepsilon _{\infty }$
as a large gap and large corrections. These corrections, in turn, are
defined by the second variation of the exchange-correlation energy, $K_{xc}(%
{\bf r,r}^{\prime })=\delta ^{2}E_{xc}/\delta n({\bf r})\delta n({\bf r}%
^{\prime }).$The LDA provides correct $K_{xc}({\bf r,r}^{\prime })$ only in
metals and only in the long range limit.

One of the first attempts to correct this was the formulation of
ADA, where $n({\bf r}^{\prime })$ in Eq. (\ref{exc}) is
substituted by $n({\bf r}),$ so that $E_{xc}^{ADA}=\int n({\bf r})\epsilon
_{xc}[\tilde{n}({\bf r)]}d{\bf r.}$ Then $\tilde{n}({\bf r)}$ is defined as $%
\tilde{n}({\bf r)=}\int w[|{\bf r-r}^{\prime }|,\tilde{n}({\bf r})]n({\bf r}%
^{\prime }{\bf )}d{\bf r}^{\prime },$ and the universal function $w$ is
chosen so that $\delta ^{2}E_{xc}^{ADA}/\delta n({\bf r})\delta n({\bf r}%
^{\prime })$ gives the correct $K_{xc}({\bf r,r}^{\prime })$ for the uniform
gas. Contrary to the WDA, the ADA is not self-interaction free in one
electron systems, and thus was never as popular as WDA.

From the beginning there was substantial interest in the behavior of WDA in
the delocalized limit \cite{IEG}. Williams and von Barth \cite{IEG(WB)}
suggested that the WDA should give substantial improvement over the LDA in
this limit, but till now no systematic study has been reported. If this
conjecture is true, the WDA has a great advantage over any other known
approximation to the DFT in the sense that it would accurately
 reproduce two key
physical limits. At least {\it some} improvement over LDA is to be expected:
 in
the short range $K_{xc}^{WDA}({\bf r,r}^{\prime })$ remains finite, and
correspondingly decay with $q\rightarrow \infty $ in reciprocal space.
Smaller $K_{xc}$ will result in weaker local field corrections and in an
improvement in $\varepsilon _{\infty }$. It is not clear {\it a priori},
though, how much $K_{xc}^{WDA}(q)$ is improved over the LDA over the whole $q
$ range and, if the improvement is only modest, whether or not an
approximation based on the WDA exists that does provide proper behavior. Here
 we derive an expression for $K_{xc}$ in the WDA, calculate $K_{xc}
$ for popular flavors of WDA, and discuss construction of a WDA method with
improved $K_{xc}$.

We start by deriving a closed expression for $K_{xc}$ in the WDA for
arbitrary $G.$ First some notation: denote the product $(e^{2}/r)G(r)$ as $%
W(r),$ use atomic units ($e=1,$ $\hbar =1),$ and use primes for the
derivative with respect to the density argument, {\it e.g.} $G^{\prime
}=dG/dn.$ We also introduce two functions, reflecting implicit dependence of
the weighted density $\bar{n}$ on variations of the real density: 
\begin{eqnarray}
d({\bf r}^{\prime }{\bf -r}) &=&\delta \bar{n}({\bf r}^{\prime })/\delta n(%
{\bf r}) \\
f({\bf r}^{\prime }{\bf -r,r}^{\prime }{\bf -r}^{\prime \prime }) &=&\frac{%
\delta ^{2}\bar{n}({\bf r}^{\prime })}{\delta n({\bf r})\delta n({\bf r}%
^{\prime \prime })}=\frac{\delta d({\bf r}^{\prime }{\bf -r})}{\delta n({\bf %
r}^{\prime \prime })}.
\end{eqnarray}
Using the WDA expression for the exchange-correlation energy, 
\begin{equation}
E_{xc}=(1/2)\int n({\bf r})n({\bf r}^{\prime })W[|{\bf r-r}^{\prime }|,\bar{n%
}({\bf r)]}d{\bf r}d{\bf r}^{\prime },  \label{EWDA}
\end{equation}
we  express $K_{xc}$ in terms of functions $d$ and $f,$ and find
these functions using the normalization condition. 
\begin{equation}
\int d{\bf r}^{\prime }n({\bf r}^{\prime })G[|{\bf r-r}^{\prime }|,\bar{n}(%
{\bf r})]=-1.  \label{norm}
\end{equation}
We proceed then in reciprocal space, which corresponds to using density
perturbation of the form $\delta n({\bf r)=}n_{q}e^{i{\bf qr}}$. Let $W_{q},$
$G_{q},$ $d_{q}$ and $f_{p,q}$  be the Fourier transforms of the
corresponding functions. Then 
\begin{equation}
d_{q}=-G_{q}/ng_{0}^{\prime }.  \label{d}
\end{equation}
Since at $q\rightarrow 0$ the LDA should be restored, 
\begin{equation}
\int d{\bf r}^{\prime }W[|{\bf r-r}^{\prime }|,n]=2\epsilon _{xc}/n.
\end{equation}
From this it immediately follows that 
\begin{equation}
G_{0}=-1/n,\qquad W_{0}=2\epsilon _{xc}/n.
\end{equation}
Thus $d_{q}=-nG_{q}.$ Next variation of Eq.(\ref{norm}) gives us $f_{p,q}.$
In fact, we need only diagonal elements, $f_{q,-q}$, for which we find $%
f_{q,-q}=2nG_{q}(nG_{q}^{\prime }+G_{q}).$ The second variation of Eq.(\ref
{EWDA}) in terms of $d$ and $f$ is 
\[
K_{xc}(q)=W_{q}+nd_{q}W_{q}^{\prime }+nd_{q}W_{0}^{\prime }+\frac{n^{2}}{2}%
(d_{q}^{2}W_{0}^{\prime \prime }+n^{2}f_{q,-q}W_{0}^{\prime }).
\]
resulting in 
\begin{equation}
K_{xc}(q)=W_{q}-n^{2}G_{q}(W_{q}^{\prime }+W_{0}^{\prime
})+n^{2}(n^{2}G_{q}^{2}W_{0}^{\prime })^{\prime }/2.  \label{kxcWDA}
\end{equation}

The original formulation of the WDA used the  homogeneous
electron gas function for $G$. Since then, three forms of $G$ have been used
in   calculations, all of which result in improvement over LDA (in the
admittedly limited number of tests performed to date). These are: the
function $G$ derived for the uniform gas by Perdew and Wang\cite{PW}, the
Gunnarsson-Jones function $G^{GJ}(r)=C_{1}(n)\{1-\exp [-\left( \frac{r}{%
C_{2}(n)}\right) ^{-5}]\},$ and the Gritsenko {\it et al}. \cite{Gri}
function $G^{GRBA}(r)=C_{1}(n)\exp [-\left( \frac{r}{C_{2}(n)}\right)
^{k}]\},$ $k=1.5$ (note that the uniform gas function\cite{PW} is
approximately given by the same expression with $k=2$). We tested these
functions for the densities $r_{s}=1,$ $2,$ and 5 and obtained modest
agreement with the Monte Carlo results\cite{MC} (Cf. the left panel of
Fig.2, where we plot
the calculated exchange-correlation local field factor $I_{xc}(q)=\frac{q^{2}%
}{4\pi }K_{xc}(q),$ and compare it with Monte Carlo data\cite{MC}). By
construction, $K_{xc}(0)$ is correct (and is in fact the LDA value). At $%
q\approx 1.5-1.8k_{F}$ $K_{xc}$ falls below its LDA value and continues to
decrease at large $q$'s. However, a closer look reveals some
disagreements: first, $I_{xc}^{WDA}(q)$ is considerably larger than the
Monte-Carlo data for the wave vectors between $\approx 0.5k_{F}$ and $%
1.5k_{F}$. Second, $I_{xc}(q)$ in WDA tends to a constant value, while in
Monte Carlo calculations it is $K_{xc}(q)$ itself that has a finite limit at 
$q\rightarrow \infty ,$ and $I_{xc}(q)\rightarrow const\cdot q^{2}$ at $%
q\rightarrow \infty .$

\begin{figure}[b!]
\begin{minipage}[b]{.49\linewidth} % [b] => Ausrichtung an \caption
 \centerline{ \epsfig{file=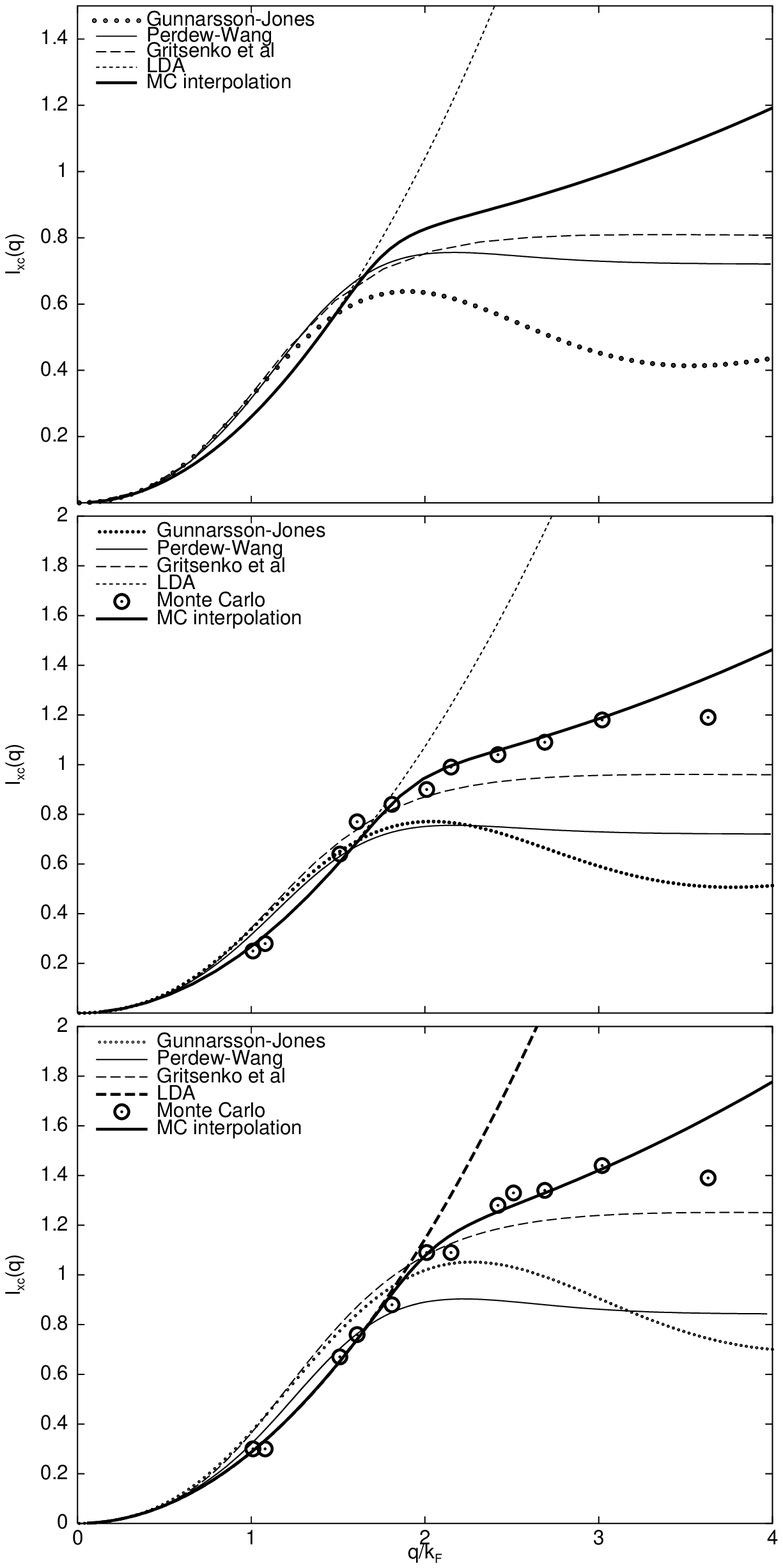,width=0.99\linewidth}}
 \vspace{0.1in}%\nopagebreak
\end{minipage}
\hspace{.01\linewidth}% Abstand zwischen Bilder
\begin{minipage}[b]{.49\linewidth} % [b] => Ausrichtung an \caption
  \centerline{\epsfig{file=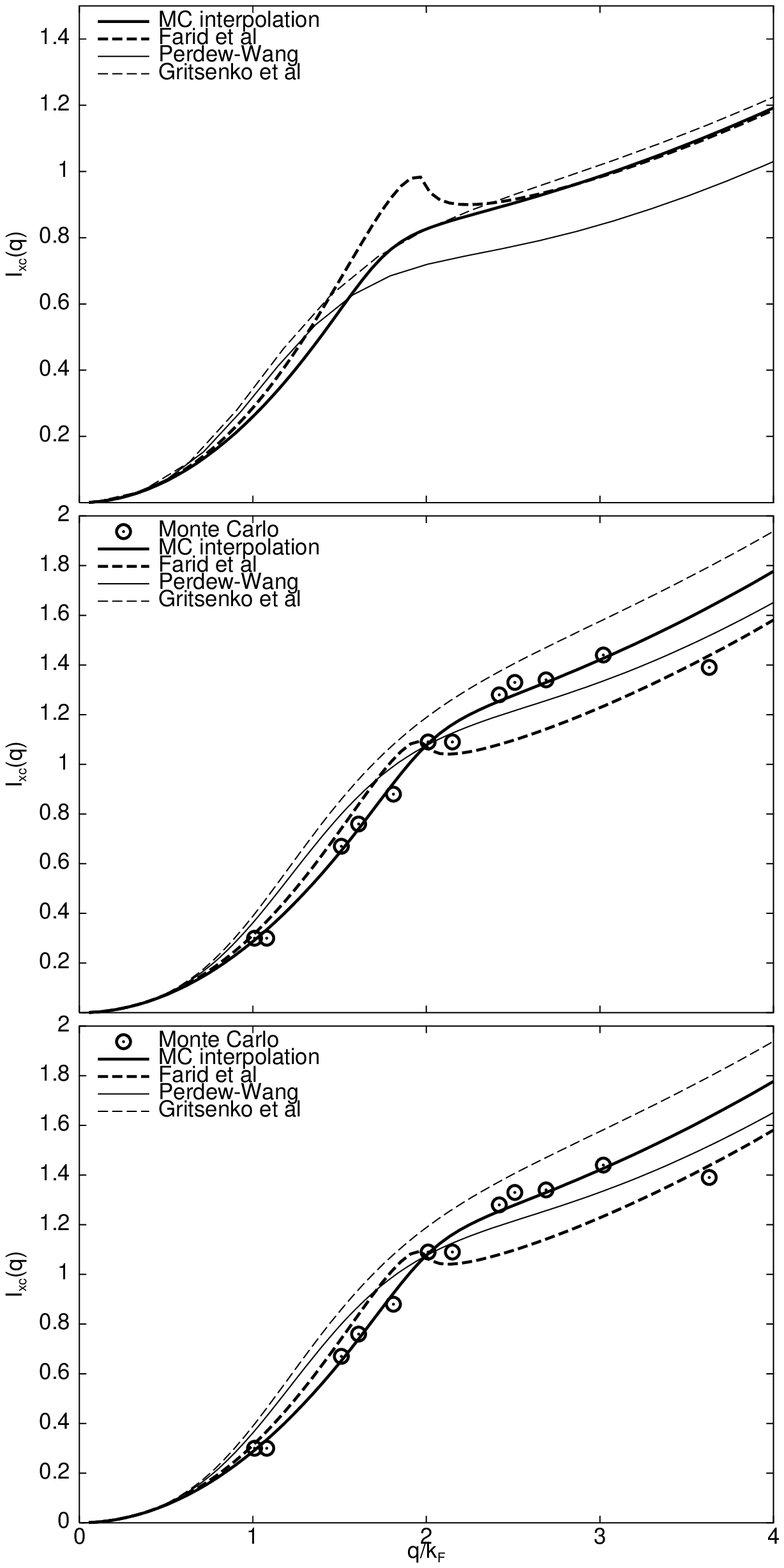,width=0.99\linewidth}} \vspace{%
0.1in}%\nopagebreak
\end{minipage}
  \caption{Left panel: 
exchange-correlation local field factor in the WDA of Ref.
\protect\onlinecite{GJ} (Gunnarsson-Jones), Ref.\protect\onlinecite{Gri}
(Gritsenko et al.),
and derived from the homogeneous electron gas pair correlation function
(Perdew-Wang), as compared with the Monte Carlo results (Monte Carlo) and
the interpolating formula thereof (MC interpolation), as given in Ref.
\protect\onlinecite{MC}.
Densities, from top to bottom, correspond to $r_s=1,2,5$.
  Right panel: the same for the modified WDA of
 Eq.\ref{awda}. Also the analytical formula of Farid
{\it et al.} Ref. \protect\onlinecite{farid} is shown.}
\end{figure}

Can one correct these two deficiencies without compromising the correct
one-electron limit of WDA? As discussed above, there is no particular reason
to use the homogeneous electron gas pair correlation function for $G$ (nor,
as discussed above, the exact pair correlation function for the
inhomogeneous system, even if it had been known). Since using $G_{0}$ in WDA
does not guarantee any improvement in describing properties of the
homogeneous gas itself, one may use the freedom in $G(r)$ to adjust the WDA
so that the calculated local field factor (and thus linear response
function) is as accurate as possible. Inversion of eq.(\ref{kxcWDA}) yields $%
G(q)$ for a given $K_{xc}(q).$ It does not guarantee, however, that the
result will be physical. So, as a first step, let us analyze Eq. (\ref
{kxcWDA}). Let us first mention that the real  $K_{xc}(q)$ changes its
behavior from the long range limit to the short range limit near $q=2k_{F},$
which plays the role of the inverse length scale. What is the characteristic
length scale in WDA? To find that, we write $G_{q}=-\varphi (p/Q)/n$, with
the condition $\varphi (0)=1,$ where $Q$ is some constant (both the
Gunnarsson-Jones and the Gritsenko {\it et al.} functions are of this form).
Then 
\begin{equation}
W_{0}=\frac{2}{\pi }\int_{0}^{\infty }G_{q}dq=-\frac{2Q}{\pi n_{0}}%
\int_{0}^{\infty }\varphi (x)dx=\frac{2\epsilon _{xc}(n_{0})}{n_{0}}. 
\nonumber
\end{equation}

If we now define $Q(n)=-\pi \epsilon _{xc}(n),$ then the second condition on 
$\varphi (x)$ becomes $\int_{0}^{\infty }\varphi (x)dx=1.$ These two
conditions reduce our freedom to adjust $G_{q}:$ since the characteristic
size of $\varphi (x)$ is of order of 1, the wave vector dependence of $G_{q}$
is defined by the ratio $q/Q=-q/\pi \epsilon _{xc}.$ Apparantely, varying
the shape of the function $\varphi $ will not significatntly change the
lentgh scale of the resulting  $K_{xc}(q).$ Explicitly density-dependent
functions $\varphi (x,n)$ may be needed to shift the hump from its
position of $q\approx 1.5k_{F}$ to $q\approx 2k_{F}.$ It is still an open
question whether or not a physically sound function can be found with this
property.

However, even if the ``$2k_{F}$'' problem is fixed, another, probably even
more important problem remains: the short wave length behavior of $K_{xc}.$
It is easy to see that if $G(q)\rightarrow 0$ at $q\rightarrow
\infty ,$ then $W_{p}\rightarrow const/p^{2}$ at $p\rightarrow \infty ,$ and
so does, according to Eq.(\ref{kxcWDA}), $K_{xc}.$ On the other hand, as
mentioned above, the correct $K_{xc}(q)$ goes to a constant
 at $q\rightarrow \infty $
as $q^{2},$ although the constant is smaller than $K_{xc}^{LDA}$.
This result was predicted by Holas \cite{Holas} and is
physically important: it comes from the exchange-correlation contribution to
kinetic energy (which is essentially local and decays slower with $q$ than
the interaction part of $E_{xc}).$ The present WDA misses the corresponding
physics. Fortunately, this is easy to correct. Farid {\it et al.}\cite{farid}
tabulated the coefficient $\gamma $ that defines the asymptotic behavior of $%
K_{xc}(q\rightarrow \infty )$ as $K_{xc}(q\rightarrow \infty )=-\frac{4\pi }{%
q^{2}}\gamma (n)\frac{q^{2}}{k_{F}^{2}},$ where $\gamma (n)$ is a universal
function, parametrized in Ref.\cite{farid}. Let us now modify the function $%
G(r)$%
\begin{equation}
G(r)=G_{1}(r)+G_{2}(r)=A\delta (r)/4\pi r+G_{2}(r),  \label{f1}
\end{equation}
Since $\int G_{1}(r)r^{2}dr=0,$ the normalization condition for $G_{2}$ is
the same as for $G$ itself. Since $4\pi \int G_{1}(r)rdr=A,$ the LDA limit
condition for $G_{2}$ becomes 
\begin{eqnarray}
4\pi \int G_{2}(r)rdr &=&2\tilde{\epsilon}_{xc}(n)/n,  \nonumber \\
\tilde{\epsilon}_{xc}(n) &=&\epsilon _{xc}(n)-An/2.  \label{f2}
\end{eqnarray}
Thus 
\begin{eqnarray}
A &=&-4\pi \gamma (n)/k_{F}^{2},  \nonumber \\
\tilde{\epsilon}_{xc}(n) &=&\epsilon _{xc}(n)+2\pi n\gamma (n)/k_{F}^{2}
\label{f3}
\end{eqnarray}
Now $G_{p}=G_{2p},$ $W_{p}=A+W_{2p},$ and $W_{p}^{\prime }=A^{\prime
}+W_{2p}^{\prime },$ %\end{multicols}
\begin{eqnarray}
K_{xc}(q)&=&A+W_{2q}-n^{2}G_{2q}(A^{\prime }+W_{2q}^{\prime }+W_{2,0}^{\prime
})+n^{2}(n^{2}G_{q}^{2}W_{2,0}^{\prime })^{\prime}\nonumber\\
&=&A-n_{0}^{2}G_{2p}A^{\prime }+\tilde{K}_{xc},  \label{Kxcd}
\end{eqnarray}
where $\tilde{K}_{xc}$ is calculated from $\tilde{\epsilon}_{xc}$ in exactly
the same way as $K_{xc}$ is calculated from $\epsilon _{xc}.$ The
corresponding functional for the exchange-correlation energy is 
\begin{equation}
E_{xc}^{AWDA}=\frac{1}{2}\int \frac{n({\bf r})n({\bf r}^{\prime })}{(|{\bf %
r-r}^{\prime }|)}G[|{\bf r-r}^{\prime }|,\bar{n}({\bf r)]}d{\bf r}d{\bf r}%
^{\prime }+\int n({\bf r})\tilde{\epsilon}_{xc}[n({\bf r})]d{\bf r}.
\label{awda}
\end{equation}
%\begin{multicols}{2}
Here $G(r)$ is normalized to $\tilde{\epsilon}_{xc}(\bar{n})$ instead of $%
\epsilon _{xc}(\bar{n}).$ In practice, the first term gives rise to the
standard expression for the WDA potential\cite{GJL,AG}, and the second
yields two additional terms, one from the variation of $n({\bf r})$, and the
other arising from $\delta \bar{n}({\bf r^{\prime })}/\delta {n}({\bf r)}$.
Since we do {\it not} require that $G(r)=\int_{0}^{1}[g(r;\lambda ,\bar{n}%
)-1]d\lambda .,$ where $g$ corresponds to the uniform gas, but rather
consider it to be a flexible function satisfying two normalization
conditions, further improvement of the method should be possible along the
line described in the previous paragraph, namely the freedom in choosing $%
G(r)$ can be used to yield $K_{xc}$ according to Eq. (\ref{Kxcd}) close to
the linear response of the homogeneous electron gas, including correct
behavior near $q=2k_{F}$. In the right panel of Fig. 2,
we show $I_{xc}$ calculated according
to Eq. \ref{awda} with the different functional form of $G(r).$ Clearly, the
results are much better than either the LDA or ``conventional'' WDA. 

In short, we have calculated the exchange-correlation
local field function $K_{xc}$ in the WDA, and found that besides the
expected improvement over the LDA it has two major deficiencies: (1) it does
not have correct asymptotic behavior at $q\rightarrow \infty ,$ and (2) the
characteristic feature at $q\approx 2k_{F}$ is displaced towards smaller $q$%
's. The former can be easily corrected by adding a delta-function component
to $G(r),$ which results in Eq. (\ref{awda}). The latter is harder to fix,
but there are still unused degrees of freedom in the formalism which may be
used to tune the behavior near $2k_{F}.$ Intuitively (cf. Ref.\cite{IEG(WB)}%
), a method which retains exact one-electron limit of WDA, and at the same
time is accurate in the opposite limit of the nearly uniform electron gas,
seems promising for practical applications. However, tests on real materials
will be needed to determine whether or not this modification of the WDA is
advantageous in practice.

\section{Conclusions}

WDA calculations of the equilibrium volume of several oxides show that with
the electron gas form of $G$ and shell partitioning, the WDA yields much
improved volumes over the LDA and GGA. No case was found where the WDA
degrades the LDA results. Phonon frequencies and the ferroelectric
distortion in KNbO$_3$ are in good accord with LDA predictions provided that
the LDA calculations are performed at the experimental volume, which is
effectively the same as the WDA volume. The direction of the ferroelectric
soft mode is in good agreement with experiment, but all exchange correlation
functionals tested yield a distortion that is 20\%
 smaller than that reported
in the powder neutron experiment of Hewat.\cite{hewat} The implication is
that this displacement should be re-examined from an experimentally. The
above results combined with the fact that the WDA linear response of the
electron gas is in substantially better agreement with Monte Carlo data than
the LDA, and that, if desired, the approach has the flexibility to further
improve this property, is suggestive that the WDA may be a generally more
reliable method than the LDA.

We are thankful for helpful discussions with L.L. Boyer, H. Krakauer, O.
Gunnarsson and R. Resta. This work is supported by the Office of Naval
Research. Calculations were performed using DoD HPCMO facilities at NAVO and
ASC. The CHSSI DoD Planewave code was employed for some of the calculations.


\begin{references}
\bibitem{char1}  Ref.\onlinecite{char} is an exception, but note that the
LDA results of this work differ from most previous calculations.
\end{references}
\end{document}